\documentstyle[aps,prb,preprint]{revtex}

\begin{document}
\tightenlines
\include{psfig}
\title{High-frequency fluctuational conductivity in 
YBa$_{2}$Cu$_{3}$O$_{7-\delta}$}
\author{E.Silva$^{(1)}$, R.Marcon$^{(1)}$, R. Fastampa$^{(2)}$, M. 
Giura$^{(2)}$, S. Sarti$^{(2)}$}
\address{ $^{(1)}$Dipartimento di Fisica ``E.Amaldi'' and Unit\`{a}  INFM,\\
Universit\`{a} ``Roma Tre'', Via della Vasca Navale 84, 00146 Roma, 
Italy\\
$^{(2)}$Dipartimento di Fisica and Unit\`{a}  INFM,\\ Universit\`{a} ``La Sapienza'',
P.le Aldo Moro 2, 00185 Roma, Italy}
\date{August 16th, 2002} \maketitle
\vspace{0.3in}
We present measurements of the fluctuational excess conductivity at 48
GHz and 24 GHz in YBa$_{2}$Cu$_{3}$O$_{7-\delta}$ films above the transition 
temperature. The measurements depart from the gaussian 
prediction for finite-frequency fluctuational conductivity.  We focus on 
the region not too close to $T_{c}$, where 
the real part of the excess 
conductivity drops much faster than the prediction of the gaussian 
model.  We calculate the dynamic excess conductivity within a 
Ginzburg-Landau approach. In the calculation we insert a 
short-wavelength cutoff of the order of the inverse coherence 
length, in order to suppress high-momentum modes.  The excess 
conductivity of all samples measured can be described very 
well by the modified model.\\
PACS numbers: 74.40.+k, 74.76.-w, 78.70.Gq\\
\\
J. Low Temp. Phys., to be published.

\maketitle
\begin{figure}
\centerline{\psfig{file=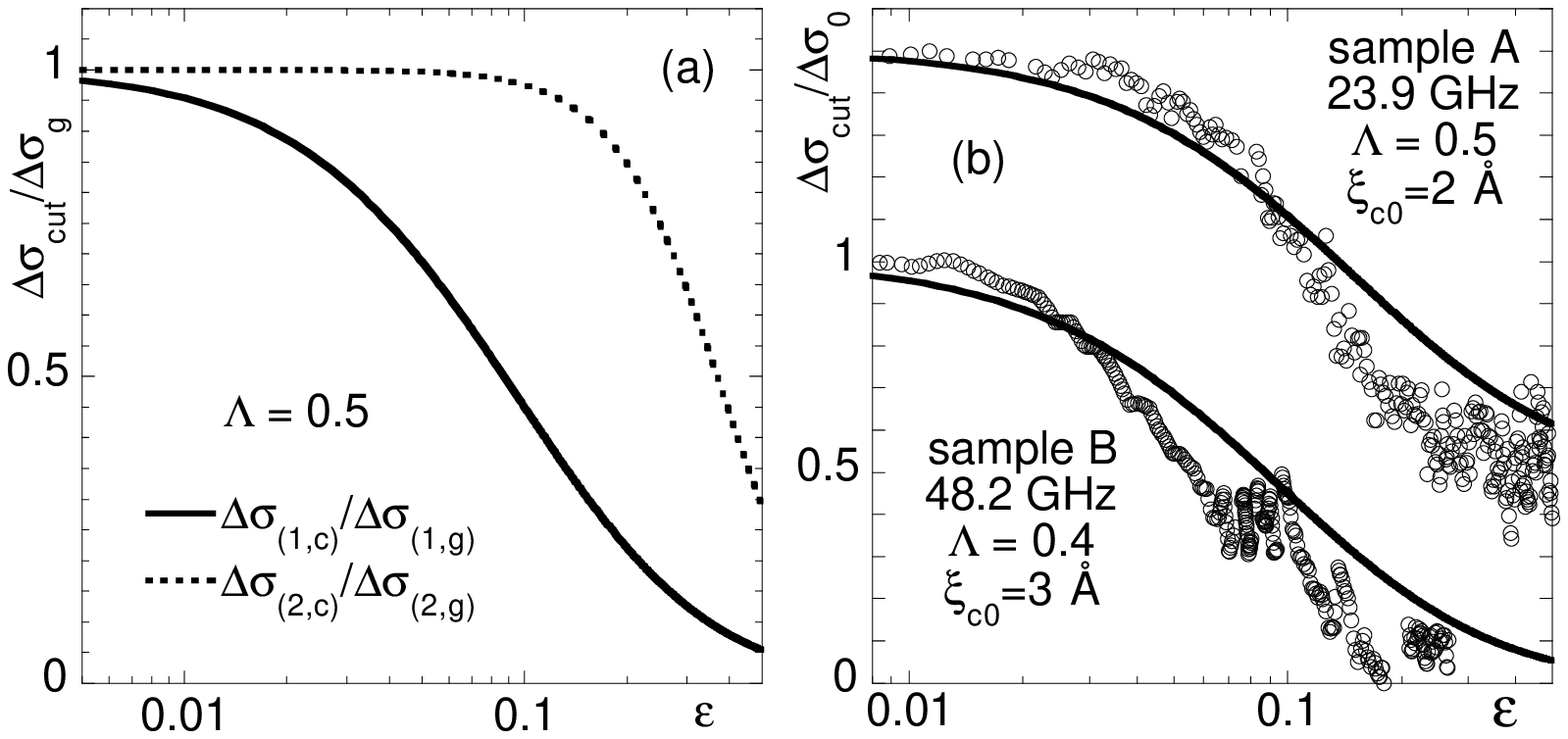,height=2.25in}}
%
\caption{(a) Ratios of $\Delta\sigma_{1,c}/\Delta\sigma_{1,g}$ 
(continuous line) and $\Delta\sigma_{2,c}/\Delta\sigma_{2,g}$ (dashed 
line) for $\Lambda$=0.5. It is seen that the cutoff correction is 
significant only on the real part. (b) Experimental ratios of 
$\Delta\sigma_{1}/\Delta\sigma_{0}$ (open symbols), together with the 
fits with the cutoffed expression.}
\label{fig1} \end{figure}
\begin{figure}
\centerline{\psfig{file=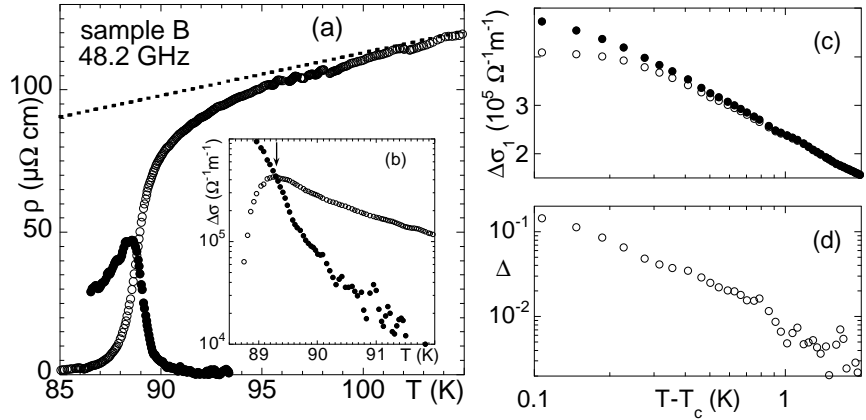,height=2.25in}}
%
\caption{Panel (a): microwave resistivity measured at 48.2 GHz in sample B. Open symbols, 
$\rho_{1}$; full dots, $\rho_{2}$; dashed line, $\rho_{n}$. Panel (b): 
extracted
complex excess conductivity.  $\Delta\sigma_{1}$, open symbols; 
$\Delta\sigma_{2}$, full dots. The arrow marks the choice of $T_{c}$. 
Panel (c): exact (open symbols) and approximate (full symbols) 
$\Delta\sigma_{1}$. Panel (d): percentage 
deviation $\Delta$ of the approximate from the exact derivation of 
$\Delta\sigma_{1}$.}
\label{fig2} \end{figure}

\vspace{0.3in}
High-$T_{c}$ superconductors are model systems for the study of superconducting 
fluctuations. High transition temperatures, short coherence 
lengths and anisotropy contribute to unveil fluctuative regimes that 
are hindered in conventional superconductors. In fact, in the latter 
the superconducting fluctuations are rather well described in terms 
of relatively simple models, such as (focussing on the electrical 
transport) the Aslamazov and 
Larkin theory\cite{aslamazov} (in a microscopic framework) or gaussian 
models.\cite{schmidt} \\
A large part of the experiments on transport properties in the 
fluctuational region have been performed by measuring the dc 
electrical conductivity.\cite{freitas,ausloos} While this approach 
benefits from the higher precision of dc measurements, it is 
intrisically unable to give information on the characteristic time of 
the fluctuations, as opposed to the dynamic conductivity.\cite{booth}  It is then 
interesting to examine appropriate tools for the analysis of the 
dynamic conductivity.\\
Most of such models have been developed to hold 
in a temperature region neither too close to $T_{c}$ (that is, outside of the 
critical region), nor too far from $T_{c}$ (where the single $q$=0
fluctuation mode is supposed to fully describe the system).  Due to 
the wide temperature region where fluctuation effects are observed in 
high-$T_{c}$ cuprates, these simple models have been recently 
reexamined.\cite{varlareview,mishonov} In this paper we 
focus on the analysis of the temperature region not too close to $T_{c}$, 
introducing a simple Ginzburg-Landau (GL)  model for the 
finite-frequency excess conductivity. We then discuss our microwave 
data for the excess conductivity in light of the proposed extension to 
the gaussian theory.\\
We briefly sketch the theoretical approach.  We derive the (a,b)-plane 
complex conductivity within the correlation function formalism, using 
the GL functional written for a three-dimensional uniaxial anisotropic 
superconductor, and limited to terms containing $\psi^{2}$.  We do not 
attempt here to include the quartic term since its effect becomes 
relevant only very close to $T_{c}$.  The dynamic of the order 
parameter, introduced through the time-dependent GL theory, is 
governed by $w=\omega\tau_{0}/\epsilon$, where $\omega/2\pi$ is the 
measuring frequency, $\epsilon$=ln$(T/T_{c})$ is the reduced 
temperature,\cite{gorkov} and $\tau_{0}$ is the 
temperature-independent GL relaxation time.\cite{skoc} The complex 
conductivity is obtained by averaging the current operator with 
respect to the noise as $\left\langle J_{ab}\left( t\right) 
\right\rangle =- \frac{\hbar e^{\star}}{m_{ab}}\int \frac{d^3 
q}{\left( 2\pi \right) ^3}q_{ab} C\left[ {\bf k}={\bf q}- \frac 
{e^{\star}} {\hbar} {\bf A}\left( t\right) ;t,t\right] $, where $C$ is 
the order-parameter correlator, ${\bf A}$ is the harmonic vector 
potential, $e^{\star}=2e$ is twice the electronic charge, and $m_{ab}$ 
is the in-plane mass of the pair.  Since\cite{booth} $\tau_{0}\sim$20 
fs, a small-$w$ expansion is appropriate.  The common approach involves integration of the 
expression for the current over all $q$ values, since the $q$=0 mode 
is the most diverging close to $T_{c}$.  The result is the well-known 
gaussian result by Schmidt:\cite{schmidt}
\begin{equation}
\Delta\sigma_{g}=\Delta\sigma_{1,g}+{\mathrm i}\Delta\sigma_{2,g}\simeq \Delta\sigma_{0}
 	 \left[\left(1-\frac{w^{2}}{16}\right)+{\mathrm i}\frac{w}{6}\right]\\
\label{sigmaapproxnocut} 
\end{equation}
\noindent
where $\Delta\sigma_{0}=\frac{e^{2}}{32\hbar \xi_c(0)\epsilon^{1/2}}$ is 
the gaussian, dc excess conductivity, and $\xi_{c}(0)$ is the zero 
temperature out-of-plane GL coherence length.
 When temperature raises above $T_{c}$, the relative weight of the 
 $q\neq$ 0 modes increases.  However modes with $q$ higher than, e.g., 
 the inverse coherence length should not be allowed.  The customary 
 way to take into account this feature is to put a cutoff in the 
 integration of the current operator (see an extended discussion in 
 Ref.\cite{mishonov}).  Using a single cutoff such that 
 $\sqrt{\sum\limits_{j=a,b,c}[q_j^{max}\xi_j(0)]^{2}}<\Lambda$, the 
 dynamic (cutoffed) excess conductivity becomes (for small 
 $w$):\cite{silvaEPJB}

\begin{eqnarray}
 \Delta\sigma_{1,c}\simeq \Delta\sigma_{0} \left\{ 
 \frac{2}{\pi}\left[atnK-\frac{K}{\left(1+K^{2}\right)^{2}}\left(1+\frac{5}{3}K^{2}\right)\right]+ 
 o(w^{2})\right\}
 \label{sigma1approx} 
 \end{eqnarray}
 
 \begin{eqnarray}
 \Delta\sigma_{2,c}\simeq \Delta\sigma_{0}  \frac{w}{6} \frac{2}{\pi} 
 \left[atnK+\frac{K}{\left(1+K^{2}\right)^{3}}\left(K^{4}-\frac{8}{3}K^{2}-1\right)\right]
 \label{sigma2approx} 
 \end{eqnarray}
 \noindent
where $K = \Lambda/\epsilon^{\frac{1}{2}}$. We note that the 
frequency dependence remains unaltered with respect to the gaussian 
expression, while the temperature dependence changes due to the cutoff 
effects. In Fig.\ref{fig1}a we plot a simulation of 
$\Delta\sigma_{1,c}/\Delta\sigma_{1,g}$ and 
$\Delta\sigma_{2,c}/\Delta\sigma_{2,g}$ as a function of $\epsilon$, 
for cutoff number $\Lambda$=0.5, and neglecting all but
the leading term in the expressions for $\Delta\sigma$. It is 
immediately seen that the cutoff effects, if present, are best visible 
in $\Delta\sigma_{1}$.
\\
We have measured the microwave response at 23.9 and 48.2 GHz in 
two YBa$_{2}$Cu$_{3}$O$_{7-\delta}$ films, grown by different 
techniques. In all cases the film was the end wall of a cylindrical metal 
cavity, resonating in the TE$_{011}$ mode. Details of the technique 
have been given elsewhere.\cite{fastampaMST} Measurements of the 
quality factor yielded the effective surface resistance $R$. The measured 
films were thinner than 2500 \AA, so that in the transition region 
$R\simeq\rho_{1}/d$, where $\rho_{1}$ is the real part of the 
resistivity and $d$ is the film thickness.\cite{silvaSUST}\\
In one sample (measured at 48.2 GHz)
we have carefully measured the frequency shift near the transition, in 
order to obtain the effective surface reactance $X\simeq\rho_{2}/d$, 
where $\rho_{2}/d$ is the imaginary part of the resistivity. The data 
are reported in Fig.\ref{fig2}a. We 
discuss with some length these data, in order to illustrate the main 
aspects of the analysis.\\
The complex excess conductivity $\Delta\sigma$ is defined as follows:
\begin{equation}
\rho_{1}+{\mathrm i}\rho_{2}=
\frac{1}{\sigma_{n} + \Delta\sigma}
\label{rho} 
\end{equation}
\noindent where $\rho_{n}=\frac{1}{\sigma_{n}}$ is the normal state 
resistivity, that we take as the linear extrapolation from high 
temperatures.  The inversion of Eq.\ref{rho} gives $\Delta\sigma_{1}$ 
and $\Delta\sigma_{2}$, that are reported as a function of $T$ in 
Fig.\ref{fig2}b.  Such data are first used to evaluate $T_{c}$, using 
the scaling theories\cite{dorsey} prediction 
$\Delta\sigma_{1}(T_{c})=\Delta\sigma_{2}(T_{c})$. 
Since $\Delta\sigma_{2}$ is vanishingly small 
$\sim$ 1 K above $T_{c}$, we can approximate 
$\Delta\sigma_{1}\simeq\frac{1}{\rho_{1}}-\frac{1}{\rho_{n}}$ not too 
close to $T_{c}$. The result is reported in Fig.\ref{fig2}c, and 
compared to the exact inversion of the $\rho$ data. As can be seen, 
above 1.01 $T_{c}$ the two procedures yield the same result. This 
feature can be put on quantitative grounds by plotting the percentage 
deviation of the approximate from the exact derivation of 
$\Delta\sigma_{1}$, as in Fig.\ref{fig2}d. It is seen that the 
deviation is largely below 5\% for $T-T_{c}>$0.5 K.\\
From this discussion of these data, we can assert that the $w\simeq$ 0 
regime for $\Delta\sigma_{1}$ is reached in a wide temperature range, 
even at our 
highest measuring frequency of 48.2 GHz.\\
On the basis of the previous discussion, we have extracted in the 
second sample $\Delta\sigma_{1}$ with the approximate expression, and 
we have used the inflection point of $\rho_{1}$ to estimate $T_{c}$ 
(we note that the precise choice of $T_{c}$ does not significantly affect the data analysis  in the
range of interest here, $\epsilon > 10^{-2}$). In Fig.\ref{fig1}b we 
report the ratio $\Delta\sigma_{1}/\Delta\sigma_{0}$, 
having used $\xi_{c}(0)$ as a scale factor in order to get the 
correct limit at 
low $\epsilon$. These data
show a clear evidence for the failure of the simple gaussian 
approximation, which predicts a constant value for all $\epsilon$. In 
the same  Fig.\ref{fig1}b, continuous lines are fits performed 
through Eq.\ref{sigma1approx}, in  good agreement with the data. We can conclude that our 
measurements of the excess conductivity, performed at microwave 
frequencies, show clear departures from the gaussian prediction, and 
that the correction is well described by the introduction of a cutoff 
in the spectrum of accessible momenta.
\\
\\
We are indebted to T.Mishonov, D.Neri,  R.Raimondi for helpful discussions.

\end{document}